\documentclass[twocolumn,amsfonts,amssymb,prl,floatfix]{revtex4-1}

\pdfoutput=1
\usepackage{graphicx}
\usepackage{bm}
\usepackage{multirow}
\usepackage{amsmath,amssymb}
\usepackage{subfigure}
\usepackage{pgf,tikz}
\usepackage[bookmarks=false]{hyperref}

\newcommand{\beq}{\begin{equation}}
\newcommand{\eeq}{\end{equation}}
\newcommand{\beqa}{\begin{eqnarray}}
\newcommand{\eeqa}{\end{eqnarray}}
\newcommand{\beqan}{\begin{eqnarray*}}
\newcommand{\eenan}{\end{eqnarray*}}

\def\Pr{{\mathrm P}}

\begin{document}
\date{2 June, 2010}

\title{Direct Entropy Determination and Application to Artificial Spin Ice}

\author{Paul~E.~Lammert$^{1}$, Xianglin~Ke$^{1}$, Jie~Li$^{1}$, Cristiano~Nisoli$^{1,2}$, 
David~M.~Garand$^{1}$, Vincent~H.~Crespi$^{1}$ and Peter~Schiffer$^{1}$}

\affiliation{
$^{1}$Department of Physics and Materials Research Institute, \\ 
{The Pennsylvania State University, 104 Davey Lab, University Park, PA 16802-6300, USA} \\
{$^{2}$CNLS and T-Division, Los Alamos National Laboratory,} Los Alamos, NM 87545, USA \\
}
\maketitle

{\bf 
From thermodynamic origins, the concept of entropy has expanded to a range of statistical measures of uncertainty, which may still be thermodynamically significant\cite{landauer61,leff-rex}.  But, laboratory measurements of entropy continue to rely on direct measurements of heat.  New technologies that can map out myriads of microscopic degrees of freedom suggest direct determination of configurational entropy by ``counting'' in systems where it is thermodynamically inaccessible, such as granular\cite{liu98,coniglio00,ohern03,corwin05,majumdar07,behringer08} and colloidal\cite{kegel00,weeks00,cui01,gasser01,alsayed05} materials, proteins\cite{pande00} and lithographically fabricated nanoscale arrays.  Here, we demonstrate a conditional probability technique to calculate entropy densities of translation-invariant states on lattices using limited configuration data on small clusters, and apply it to arrays of interacting nanoscale magnetic islands (``artificial spin-ice''\cite{wang06}).  Models for statistically disordered systems can be assessed by applying the method to relative entropy densities.  For artificial spin-ice, this analysis shows that 
nearest neighbor correlations drive longer-range ones.
}

\begin{figure}
\includegraphics[width=65 mm]{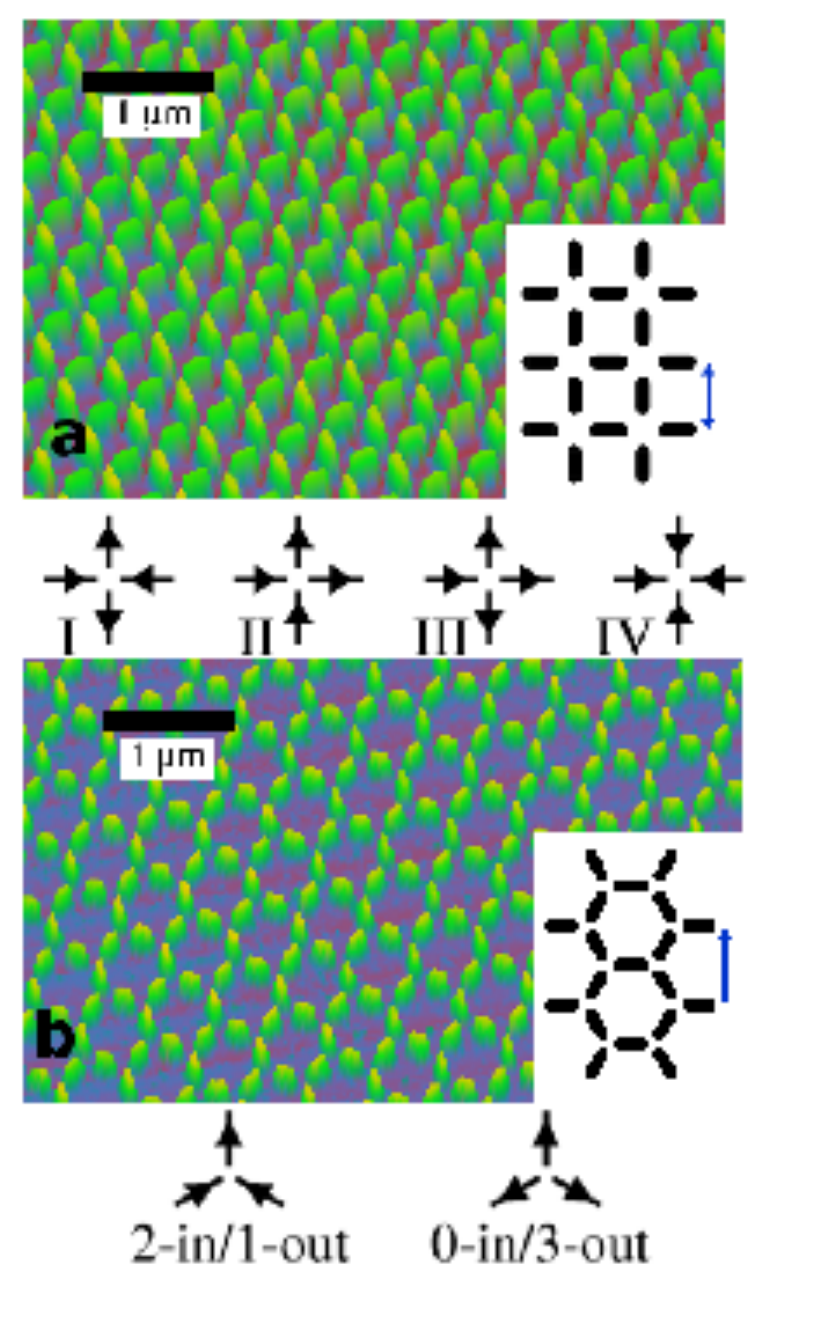}
\caption{ {\bf Artificial spin-ice arrays.}
{\bf a}.  AFM image of 400 nm square lattice with inset showing lattice spacing,
and schematics of vertex types. 
{\bf b}. Similarly for 520 nm honeycomb lattice.}
\label{photo}
\end{figure}

Our artificial spin ice\cite{wang06} systems are arrays of lithographically defined single-domain ferromagnetic islands (25 nm thick and 220 nm $\times$ 80 nm in area) on the links of square and honeycomb lattices (Fig.~\ref{photo}).  Shape anisotropy forces island moments to point along the long axes, forming effective Ising spins.  The coercive field is about 770 Oe (i.e., a barrier of order 10$^5$ K), while the field from one island on a neighbor only of order 10 Oe (10$^4$ K).  The arrays are demagnetized by rotating in an in-plane external magnetic field $H_{\mathrm{ext}}$, initially strong enough to produce complete polarization, subsequently reduced to zero in small increments\cite{wang06,ke08} $\Delta H_{\mathrm{ext}}$, with reversal of the field at each step.  For small step sizes, the result is a statistically reproducible macrostate, operationally defined by the demagnetization protocol\cite{wang06,ke08}, which is probed by magnetic force microscopy to obtain the static moments of individual islands.  We want the entropy of a single macrostate, but distinct runs might produce distinct macrostates (for example, a residual magnetization at larger step size).  In most cases, data are collected in a single run, averting the problem, but the entropy of macrostate mixture would be relatively unimportant anyway, as is shown in Supplementary Information \S3.  For large structurally regular systems such as ours, it is more appropriate to work not with total entropy, but with entropy density [See Eq. (\ref{s_as_lim})] having units of bits/island, a value of 1 corresponding to complete disorder.  

The strongest interactions, between islands meeting at a vertex, favor head-to-tail moment alignment.  But not all these can be satisfied simultaneously, resulting in a kind of frustration.  Still, for the square lattice, the ground state is only two-fold degenerate\cite{wang06}, since Type-I vertices, as defined in Fig. 1, are lowest in energy.  That the ordered ground state is never found experimentally\cite{nisoli07,ke08} suggests that the evolution is kinetically constrained\cite{ritort03,rugged}.  For instance, one spin flip converts a Type-II to a Type-III; flips of two perpendicular islands are required to reach Type-I.  In contrast to the square lattice, the honeycomb lattice has a macroscopically degenerate ground state when only nearest or next-nearest neighbor interactions are effective (longer-range interactions break the degeneracy\cite{moller09} at a much lower energy scale).  The interactions prefer a 2-in/1-out or 1-in/2-out arrangement at every vertex (``quasi-ice rule'').  This constraint alone produces a state, ideal quasi-ice, with an entropy density of 0.724 bits/island.  Interaction between (mono)pole-strengths $Q$ at neighboring vertices\cite{tanaka06,qi08} reduce the ground state degeneracy to 0.15 bits/island by favoring $Q=-1$ (2-in/1-out) next to $Q = +1$.  The contrast between the square and honeycomb lattice ground states -- two-fold degenerate versus macroscopically degenerate -- provides an opportunity to investigate the interplay between the strictures of kinetic constraint and the freedom of massive degeneracy.

\begin{figure}
\includegraphics[width=90 mm]{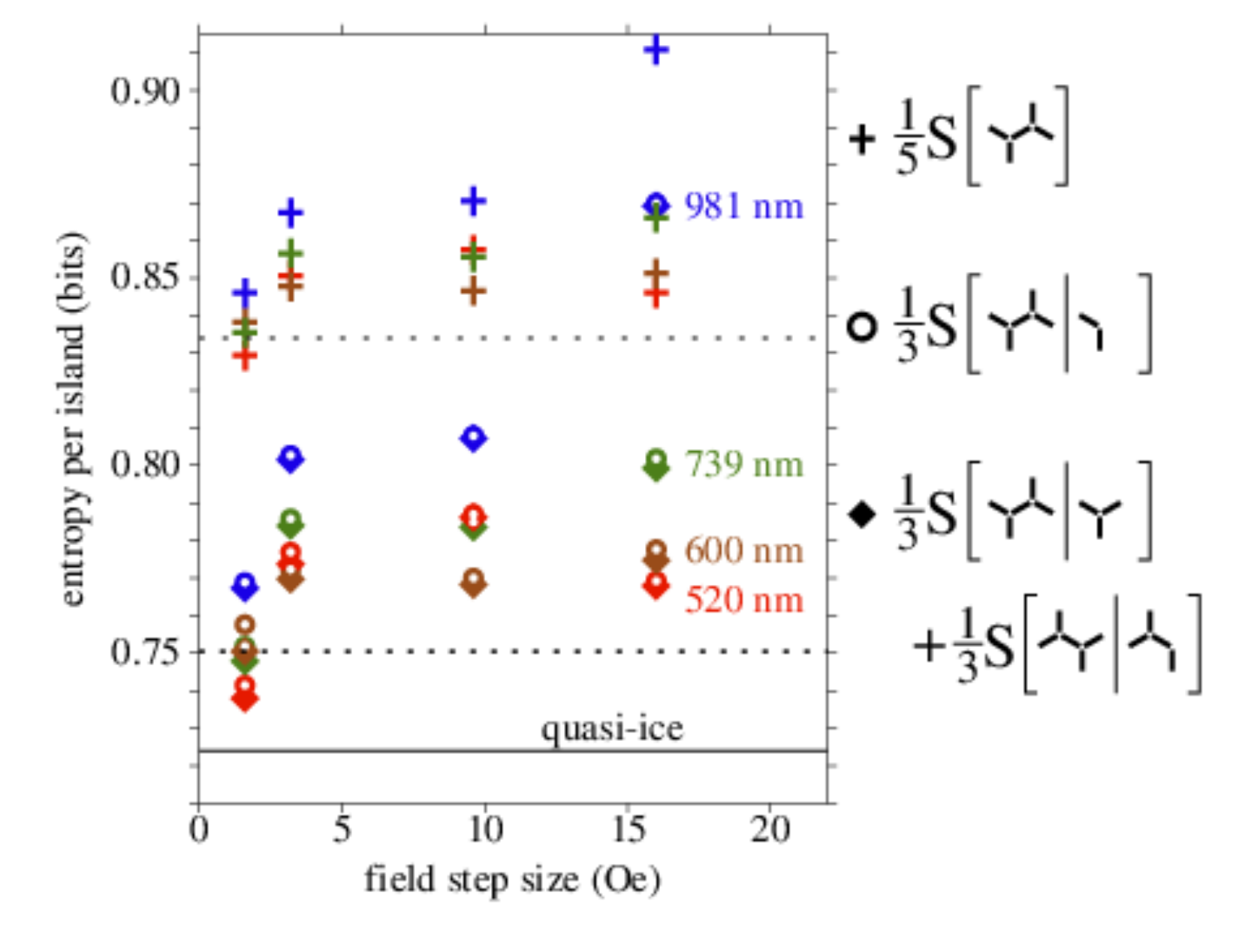}
\caption{{\bf Entropy density upper bounds for honeycomb artificial spin ice at four 
lattice constants as a function of  the demagnetization step size $\Delta H_{\mathrm{ext}}$.}
All bounds are derived using configuration statistics for the 5-island cluster $\Lambda_5$ 
shown in the inset.  Crosses are the direct estimate $S(\Lambda_5)/5$, while filled diamonds 
use inequality (\ref{2_step}) and open circles use inequality (\ref{1_step}). 
The dashed lines are the result of our technique applied to ideal quasi-ice
(every vertex Type-I with no other restrictions),
the upper from the simple cluster-estimate, and the lower two (indistinguishable) from 
the two conditioning approximations. 
The simple cluster-estimate applied to a single vertex would give 0.86 bit/island.
The solid line at 0.724 shows the actual entropy density of ideal quasi-ice.
At the smallest lattice constants and smallest step sizes, subtle signs of 
longer-ranged monopole correlations appear. 
}
\label{hex-plots}
\end{figure}

We now develop a method to extract the entropy densities on our lattices from the measured configurations of the island magnetic moments. Consider a finite cluster $\Lambda$ of islands, for example, the 5-island cluster comprising two adjacent vertices (See Fig. \ref{hex-plots} legend) and the collection of random variables $\sigma_\Lambda$ which are the spins belonging to $\Lambda$.  The Shannon(-Gibbs-Boltzmann) entropy\cite{wehrl78,penrose70,jaynes57} of $\Pr_\Lambda(\sigma_\Lambda)$, the distribution of $\sigma_\Lambda$, is
 \begin{equation}
S(\Pr_\Lambda) = -\sum_{\sigma_\Lambda} \Pr(\sigma_\Lambda) {\log}_2 \Pr(\sigma_\Lambda),
\label{shannon}
\end{equation}
where the sum runs over all possible values of the random variable(s) $\sigma_\Lambda$.  Note that $S$ is rendered dimensionless by omitting Boltzmann's constant, and the base of the logarithm is 2, so that the units are bits.  If $\Lambda$ is taken ever larger while the fraction of islands on the edge tends to zero (van Hove limit), we obtain the bulk entropy density $s$:
\begin{equation}
{s} = \lim_{\Lambda\nearrow\infty} 
\frac{S(\Pr_\Lambda)}{|\Lambda|}.
\label{s_as_lim}
\end{equation} 
If each island moment independently points either way with probability $1/2$, then the entropy density is one bit per island, the largest possible.  Lower entropy density indicates correlations in a generic sense.  For example, the fully-polarized initial state created by a large $H_{\mathrm{ext}}$ has zero entropy density.

The obvious approximation to $s$ suggested by Eq. (\ref{s_as_lim}) is simply ${S(\Pr_\Lambda)}/{|\Lambda|}$ for the biggest practicable cluster.  But this ``simple cluster-estimate'' is not very good because the configuration space grows exponentially with cluster size $|\Lambda|$, while boundary-crossing correlations are completely neglected.  To understand the latter point, suppose the entire lattice covered without gaps or overlap by translates of $\Lambda$.  The state constructed from the marginals of $\Pr$ on those translates, {\em taking them independent}, has entropy density exactly $S(P_\Lambda)/|\Lambda|$.  However, short-range boundary-crossing correlations are the same as corresponding intra-cluster correlations, so are reflected in small-cluster data and can be properly counted using {\it conditional entropy}.  The method resembles one proposed some years ago\cite{schlijper,schlijper-PRA} for Monte Carlo simulations of lattice spin models in equilibrium.

One way to think of the total entropy of a given macrostate is as the average uncertainty about the {\em particular} microstate at hand.  Imagine a microstate of the honeycomb lattice revealed three islands (one vertex) at a time, row-by-row. One instant in the process looks like this (the grey vertex is about to be revealed):
\begin{minipage}{4.0 cm}
\begin{tikzpicture}[scale=0.4]
\begin{scope}[line width=2.0pt]
\foreach \i in {1,...,4}
{
\draw[shift={(\i*1.732,-1.8)}] (90:0.15) -- (90:0.85) 
                (-30:0.15) -- (-30:0.85) (210:0.15) -- (210:0.85);
}
\foreach \i in {1,2,3}
{
\draw[shift={(\i*1.732-0.86602,-0.3)}] (90:0.15) -- (90:0.85) 
                (-30:0.15) -- (-30:0.85) (210:0.15) -- (210:0.85);
}
\draw[shift={(3.5*1.732,-0.3)},color=black!25] (90:0.15) -- (90:0.85) 
                (-30:0.15) -- (-30:0.85) (210:0.15) -- (210:0.85);

\draw (0.1,0.0) circle (0.5 pt);
\draw (-0.3,0.0) circle (0.5 pt);
\draw (-0.7,0.0) circle (0.5 pt);

\draw (7.7,-1.5) circle (0.5 pt);
\draw (8.1,-1.5) circle (0.5 pt);
\draw (8.5,-1.5) circle (0.5 pt);

\draw (3.4,-2.4) circle (0.5 pt);
\draw (3.4,-2.8) circle (0.5 pt);
\draw (3.4,-3.2) circle (0.5 pt);

\draw (0.8,-1.5) circle (0.5 pt);
\draw (0.4,-1.5) circle (0.5 pt);
\draw (0.0,-1.5) circle (0.5 pt);

\end{scope}
\end{tikzpicture}
\end{minipage}
Neglecting the (far distant) lattice edge, each newly revealed vertex bears the 
same spatial relation to those already known, so the revelation, on average,
reduces the uncertainty by exactly 3 times the entropy per spin.
Cast this way, the entropy density appears as a {\it conditional\/} entropy
\cite{cover-thomas}.
The conditional entropy of $\sigma_\Lambda$ given $\sigma_\Gamma$ is
\begin{equation}
S(\sigma_\Lambda | \sigma_\Gamma) = -\sum_{\sigma_\Lambda,\sigma_\Gamma} 
\Pr(\sigma_\Lambda,\sigma_\Gamma)
\log_2 \Pr(\sigma_\Lambda| \sigma_\Gamma),
\label{conditional-def}
\end{equation}
where $\Pr(\sigma_\Lambda|\sigma_\Gamma) = \Pr(\sigma_{\Lambda},\sigma_\Gamma)/\Pr(\sigma_{\Gamma})$ is the conditional probability of $\sigma_\Lambda$ given $\sigma_\Gamma$. 
The joint entropy of $\sigma_\Lambda$ and $\sigma_\Gamma$ then
has the pleasant decomposition
$S(\sigma_\Lambda,\sigma_\Gamma) = S(\sigma_\Lambda | \sigma_\Gamma) + S(\sigma_\Gamma)$.
(Learning $\sigma_\Gamma$ and $\sigma_\Lambda$ at once is the same as 
learning $\sigma_\Gamma$ and then $\sigma_\Lambda$.) 
Note that if $\Lambda$ and $\Gamma$ overlap, common spins contribute zero
to $S(\sigma_\Lambda | \sigma_\Gamma)$.

As a simple illustration, suppose $\Lambda$ and $\Gamma$ are single islands, with the probabilities for $P(\sigma_\Lambda,\sigma_\Gamma)$ being given by $\Pr(\downarrow,\uparrow) = 0$ and $\Pr(\uparrow,\uparrow) = \Pr(\uparrow,\downarrow)=\Pr(\downarrow,\downarrow) = 1/3$. If we know that $\sigma_\Gamma = \uparrow$, then the remaining uncertainty about $\sigma_\Lambda$ is zero, but if we know that $\sigma_\Gamma = \downarrow$, then the uncertainty is total: 1 bit. Weighting by the probabilities of $\sigma_\Gamma$ to be $\uparrow$ or $\downarrow$ gives the entropy of $\sigma_\Lambda$ conditioned on $\sigma_\Gamma$: $\Pr(\sigma_\Gamma=\uparrow)\cdot(0) +\Pr(\sigma_\Gamma=\downarrow)\cdot(1) = 2/3$ bit.

The Methods section explains how conditional entropy and other basic  notions of information theory can be used to obtain good approximations to the entropy density $s$ from limited data.  The result of applying two such approximations to the experimental data for honeycomb lattices are plotted in Fig. \ref{hex-plots} as a function of field step size $\Delta H_{\mathrm{ext}}$ for each lattice constant, along with one simple cluster-estimate for comparison.  Data-set sizes are reported in Supplementary Information \S1.  The simple cluster-estimate $S(\Lambda)/|\Lambda|$ using the five-island di-vertex (Fig. \ref{hex-plots} legend) provides a very poor bound compared to our conditioning technique.  Reducing lattice constant or step size should lower the entropy since the first leads to stronger interactions, and the second gives interactions a better chance to be the decisive factor for island flips.  The expected lattice constant trend is seen but there is an unexpected plateau with respect to $\Delta H_{\mathrm{ext}}$.

It makes sense to compare the experimental states to ideal honeycomb quasi-ice through entropy densities.  That of honeycomb quasi-ice is 0.724 bit/island (Supplementary Information \S2).  But proper comparison requires using the same estimates for both systems.  Dashed lines in the plot show the estimates for the model system, the upper for the simple cluster-estimate (compare crosses) and the lower two (indistinguishable) for the conditioning estimates.  Hence, the 520 nm array at $\Delta H_{\mathrm{ext}}=1.6$ Oe has {\it less\/} entropy than ideal quasi-ice.  This can be explained by correlations between net magnetic charge $Q=\pm 1$ of nearest-neighbor vertices.  Ideal quasi-ice has a weak anticorrelation: $\langle Q_i Q_j \rangle \approx -1/9$.  In some samples, this correlation reaches -0.25, reflected in a small entropy decrease.  Complete sublattice ordering, $\langle Q_i Q_j \rangle = -1$, would reduce the entropy 
to $s\approx 0.15$ bit/island (Supplementary Information \S2).  This extra correlation may explain the slightly better performance of the bound with a complete vertex in the conditioning data.

\begin{figure}
\includegraphics[width=90 mm]{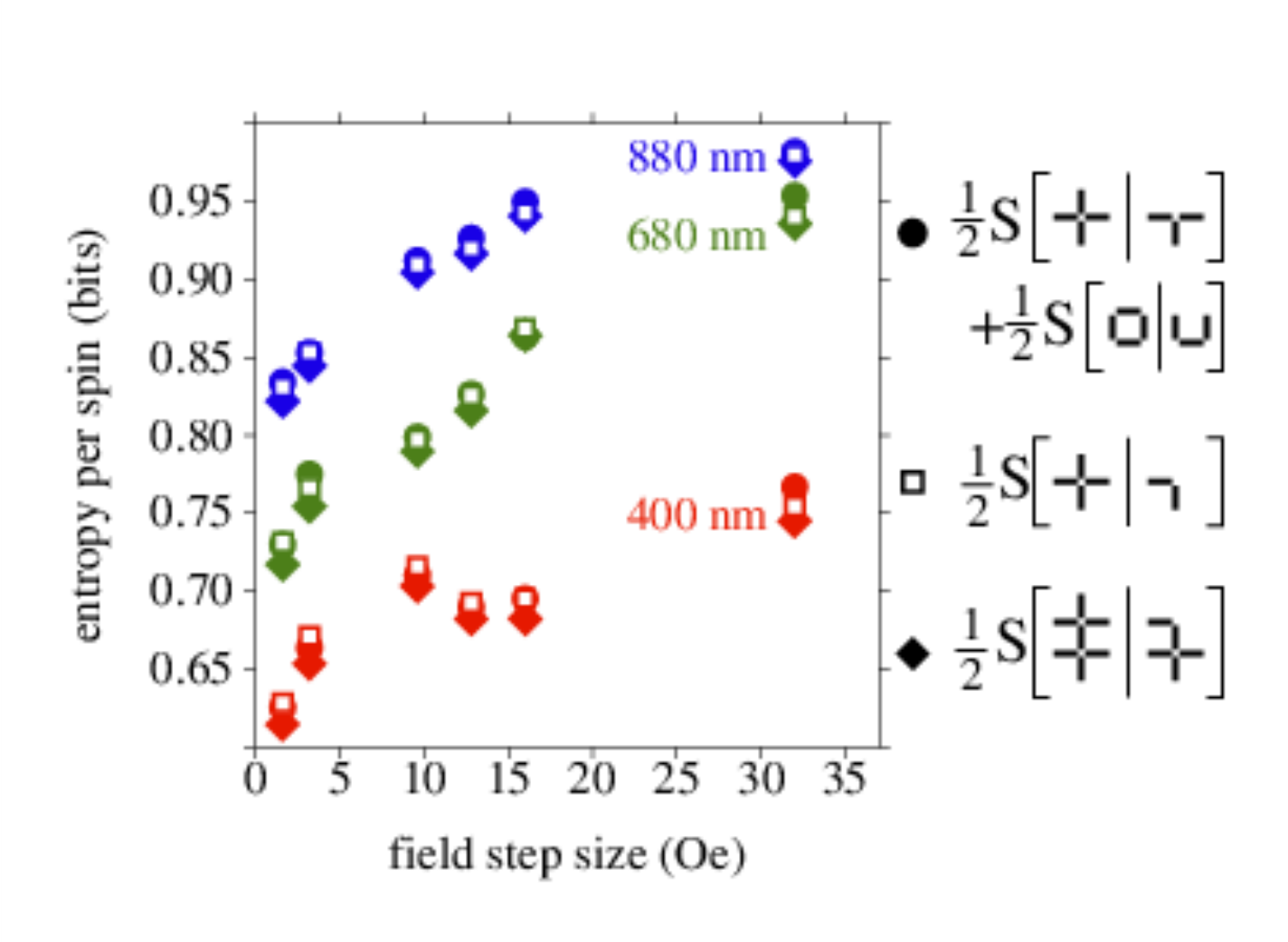}
\caption{{\bf Entropy density upper bounds for square lattice artificial spin ice.}
The three approximations agree closely.  Even extrapolated to zero step size, 
AC demagnetized square spin ice never approaches the 
ground state.  The diamond estimate is always lower than the square because the former 
retains more conditioning, whereas the added islands are the same. 
Dataset sizes are reported in Supplementary Information.  }
\label{square-plots}
\end{figure}

The entropy of honeycomb artificial spin ice reveals a state close to ideal quasi-ice, with slight antiferromagnetic vertex charge ordering.  The contrasting failure of AC demagnetization of the square lattice to approach the completely ordered ground state is precisely quantified by entropy.  We use three approximations (upper bounds) for the square lattice entropy density.  They are found by a procedure parallel to that for the honeycomb lattice and are shown in the legend of Fig. \ref{square-plots}.  
The three agree well, and this rough convergence test suggests that they are close to the true entropy densities.  As for the honeycomb lattice, we expect smaller entropy for smaller $\Delta H_{\mathrm{ext}}$ or smaller lattice spacing.  In general, this seems to be the case, but the ground state is never approached.  Even extrapolations $\Delta H_{\mathrm{ext}} \rightarrow 0$ have large entropy densities.

Closer inspection suggests jamming at $\Delta H_{\mathrm{ext}}$.
Kinetically constrained approach to ground states defines behavior of complex systems across many fields\cite{rugged}, such as protein folding\cite{pande00}, self-assembly, glasses and granular systems\cite{liu98,coniglio00,ohern03,corwin05,majumdar07,behringer08}.  Ergodicity is thwarted by both tall energy barriers and configuration space constrictions, combined into free energy barriers.  An ergodic system explores all of configuration space, whereas folding proteins live within a ``folding funnel.'' This dynamic constriction of allowed configurations introduces many deep conceptual challenges. AC demagnetized artificial spin ice puts the conceptual challenge of kinetic constraint into sharp relief: as the rotating external field weakens, islands one-by-one fall out of ``field-following'' mode, driven by inter-island interactions that suppress the local depolarization field and lock in the orientation of the fallen-away island. Thus each spin likely makes only a single decision on how to point upon escaping coercion, with no prospect of later surmounting barriers. The system's approach to the ground state is essentially one-way. Thus it is not surprising that only a macroscopically degenerate ground state target can be ``hit.'' Notwithstanding this failure of ergodicity, square-lattice artificial spin ice can still be described by a statistical model based, like thermodynamics, on maximum entropy\cite{nisoli07,ke08}. As an extreme case of kinetic constraint, rotationally annealed artificial spin ice can afford unique insights into statistical mechanics of complex systems. For example, array topology can control the ground-state degeneracy, as seen here.

Even without detailed knowledge of how the final square lattice states develop, a concisely descriptive model may be sought.  We conjecture that the lattice state is fully determined by {\em correlations} between nearest-neighbor pairs diagonally, or straight, across a vertex, and thus model it by a constrained maximum entropy state, which is as random as possible, consistent with those correlations.  Adapting the conditioning techniques (see Methods section), we can efficiently estimate the entropy density of the experimental states relative to the maximum entropy state, $s(\mathrm{expt}||\mathrm{ME})$.  This global measure of dissimilarity does not depend on identifying the ``right'' correlations, and allows an assessment of the model.  Results are given in the following table.

%\begin{table}
%\break\vspace*{12pt}
\medskip
\centerline{$s(\mathrm{expt}||\mathrm{ME})$ (10$^{-3}$ bit/island)}
%\smallskip
\centerline{
\begin{tabular}{c | c c c c c}
\hline
lattice (nm) & \multicolumn{5}{c}{$\Delta H_{\mathrm{ext}}$ (Oe)} \\
& 1.6 & 3.2 & 9.6 & 12.8 & 16  \\ 
\hline
% \hline
400 & 5.1 & 4.2 & 13.5 & 2.9 & 9.2 \\
% \hline
680 & 4.5 & 5.6 & 9.1 & 5.4 & 5.0 \\
% \hline
880 & 4.4 & 16.1 & 7.9 & 1.4 & 2.6 \\
\hline
\end{tabular}
}
\medskip
%\end{table}
%\break\vspace*{12pt}
Note that\cite{dembo-zeitouni} the probability that a typical experimental state in region $\Lambda$ is likely to be mistaken for a maximum entropy state decays asymptotically as $\exp[-|\Lambda| s(\mathrm{expt}||\mathrm{ME})]$.  Apparently, the entropy reduction below independent islands in the square lattices is well accounted for by nearest-neighbor correlations, and those they entail.

The reduction in the entropy of an interacting system below that of uncoupled degrees of freedom is due mostly to short-range correlations, even near a critical point. Thus, the efficient conditional entropy technique described here can be applied to a wide variety of resolvable complex systems such as granular media and colloidal systems which can now be spatially resolved in the required detail~\cite{kegel00,weeks00,cui01,gasser01,alsayed05}.  Entropy density is a general measure of order which is not tied to pre-identified correlations. Hence it is especially valuable for states such as square ice or other jammed, glassy states which are far from identifiable landmarks. 

\section{Methods}
According to the discussion around Eq. (\ref{conditional-def}),
the  entropy density $s$ of the infinite honeycomb lattice is given by 
(ignore the color for now)
\begin{equation}
\begin{tikzpicture}[scale=0.28]
\begin{scope}[line width=1.0pt]
\foreach \i in {1,3,4}
{
\draw[shift={(\i*1.732,0)}] (90:0.15) -- (90:0.85) 
                (-30:0.15) -- (-30:0.85) (210:0.15) -- (210:0.85);
}
\draw[shift={(2*1.732,0)}] (-30:0.15) -- (-30:0.85) (210:0.15) -- (210:0.85);
\draw[color=red,shift={(2*1.732,0)}] (90:0.15) -- (90:0.85);
\draw[shift={(1*1.732-0.86602,1.5)}] (90:0.15) -- (90:0.85) 
                (-30:0.15) -- (-30:0.85) (210:0.15) -- (210:0.85);
\draw[shift={(2*1.732-0.86602,1.5)}] (90:0.15) -- (90:0.85) 
                                     (210:0.15) -- (210:0.85);
\draw[color=red,shift={(2*1.732-0.86602,1.5)}] (-30:0.15) -- (-30:0.85);

\draw[color=red,shift={(3*1.732-0.86602,1.5)}] (90:0.15) -- (90:0.85) 
                (-30:0.15) -- (-30:0.85) (210:0.15) -- (210:0.85);

\draw (0.2,1.8) circle (0.5 pt);
\draw (-0.2,1.8) circle (0.5 pt);
\draw (-0.6,1.8) circle (0.5 pt);

\draw (7.6,0.4) circle (0.5 pt);
\draw (8.0,0.4) circle (0.5 pt);
\draw (8.4,0.4) circle (0.5 pt);

\draw (3.4,-0.7) circle (0.5 pt);
\draw (3.4,-1.1) circle (0.5 pt);
\draw (3.4,-1.5) circle (0.5 pt);
%% condition
\foreach \i in {1,3,4}
{
\draw[shift={(10.5 + \i*1.732,0)}] (90:0.15) -- (90:0.85) 
                (-30:0.15) -- (-30:0.85) (210:0.15) -- (210:0.85);
}
\draw[shift={(10.5 + 2*1.732,0)}] (-30:0.15) -- (-30:0.85) (210:0.15) -- (210:0.85);
\draw[color=red,shift={(10.5 + 2*1.732,0)}] (90:0.15) -- (90:0.85);
\draw[shift={(10.5 + 1*1.732-0.86602,1.5)}] (90:0.15) -- (90:0.85) 
                (-30:0.15) -- (-30:0.85) (210:0.15) -- (210:0.85);
\draw[shift={(10.5 + 2*1.732-0.86602,1.5)}] (90:0.15) -- (90:0.85) 
                                            (210:0.15) -- (210:0.85);
\draw[color=red,shift={(10.5 + 2*1.732-0.86602,1.5)}] (-30:0.15) -- (-30:0.85);

\draw (10.7,1.8) circle (0.5 pt);
\draw (10.3,1.8) circle (0.5 pt);
\draw (9.9,1.8) circle (0.5 pt);

\draw (18.1,0.4) circle (0.5 pt);
\draw (18.5,0.4) circle (0.5 pt);
\draw (18.9,0.4) circle (0.5 pt);

\draw (13.9,-0.7) circle (0.5 pt);
\draw (13.9,-1.1) circle (0.5 pt);
\draw (13.9,-1.5) circle (0.5 pt);

\end{scope}
\draw (-0.8,3.0) -- (-1.5,3.0) -- (-1.5,-2.0) -- (-0.8,-2.0);
\draw (9,3.0) -- (9,-2.0);
\draw (19,3.0) -- (19.7,3.0) -- (19.7,-2.0) -- (19,-2.0);
\draw (-4.4,0.4) node {\Large $3s=S$};
\end{tikzpicture}
\label{Cndtnl_Eq}
\end{equation}
Now we find small-cluster approximations to this entropy density, using two principles\cite{cover-thomas}. 
(A) If $\sigma_\Gamma$ is known there is no uncertainty about it, 
so $S(\sigma_\Lambda,\sigma_\Gamma|\sigma_\Gamma) = S(\sigma_\Lambda|\sigma_\Gamma)$ 
for arbitrary $\Lambda$ and $\Gamma$.
Thus, in pictorial equations, visual perspicuity will dictate retention or omission 
of conditioning variables on the left of the bar.
(B) Providing more conditioning information lessens uncertainty: 
$S(\sigma_\Lambda|\sigma_{\Gamma}) \ge S(\sigma_\Lambda|\sigma_\Gamma,\sigma_{\Gamma^\prime})$. 
Unlike a simple application of Eqn. (\ref{s_as_lim}), our conditional entropy method 
fully accounts for short-range correlations without boundary error.  
Like the simple cluster-estimate, it provides upper bounds on the true entropy density. 
Dropping all but the red islands in Eqn. (\ref{Cndtnl_Eq}) yields 
\begin{equation}
\begin{tikzpicture}[scale=0.30]
\begin{scope}[style=thick]
\draw (-90:0.15) -- (-90:0.85) (150:0.15) -- (150:0.85);
\draw[shift={(0.5*1.732,0.5)}] (90:0.15) -- (90:0.85) 
                               (-30:0.15) -- (-30:0.85) (210:0.15) -- (210:0.85);
%% condition
\draw[shift={(4,0)}] (-90:0.15) -- (-90:0.85) (150:0.15) -- (150:0.85);
%\end{scope}
\draw (-0.8,2.0) -- (-1.5,2.0) -- (-1.5,-1.6) -- (-0.8,-1.6);
\draw (2.4,2.0) -- (2.4,-1.6);
\draw (4.5,2.0) -- (5.2,2.0) -- (5.2,-1.6) -- (4.5,-1.6);
%\draw (-4.7,0) node {\Large 3s $\le$ S};
\draw (-4.7,0) node {\Large $3s \le S$};
\end{scope}
\end{tikzpicture}
\label{1_step}
\end{equation}
Alternatively, if we add a vertex in two stages rather than all at once as in Eq. (\ref{Cndtnl_Eq}), we immediately arrive at 
\begin{equation}
\begin{tikzpicture}[scale=0.25]
\begin{scope}[line width=1.0pt]
\draw[shift={(1*1.732,0)}] (-30:0.15) -- (-30:0.85) (210:0.15) -- (210:0.85);
\draw[color=red,shift={(1*1.732,0)}] (90:0.15) -- (90:0.85);
\foreach \i in {2,3}
{
\draw[shift={(\i*1.732,0)}] (90:0.15) -- (90:0.85) 
                (-30:0.15) -- (-30:0.85) (210:0.15) -- (210:0.85);
}
\draw[shift={(1*1.732-0.86602,1.5)}] (90:0.15) -- (90:0.85) 
                                     (210:0.15) -- (210:0.85);
\draw[color=red,shift={(1*1.732-0.86602,1.5)}] (-30:0.15) -- (-30:0.85);
\draw[color=red,shift={(2*1.732-0.86602,1.5)}] (90:0.15) -- (90:0.85) 
                (-30:0.15) -- (-30:0.85) (210:0.15) -- (210:0.85);
\draw (0.2,1.8) circle (0.5 pt);
\draw (-0.2,1.8) circle (0.5 pt);
\draw (-0.6,1.8) circle (0.5 pt);

\draw (6.0,0.4) circle (0.5 pt);
\draw (6.4,0.4) circle (0.5 pt);
\draw (6.8,0.4) circle (0.5 pt);

\draw (3.4,-0.7) circle (0.5 pt);
\draw (3.4,-1.1) circle (0.5 pt);
\draw (3.4,-1.5) circle (0.5 pt);
%% condition
\begin{scope}[shift={(9,0)}]
\draw[shift={(1*1.732,0)}] 
                (-30:0.15) -- (-30:0.85) (210:0.15) -- (210:0.85);
\draw[color=red,shift={(1*1.732,0)}] (90:0.15) -- (90:0.85);
\foreach \i in {2,3}
{
\draw[shift={(\i*1.732,0)}] (90:0.15) -- (90:0.85) 
                (-30:0.15) -- (-30:0.85) (210:0.15) -- (210:0.85);
}

\draw[shift={(0.86602,1.5)}] (90:0.15) -- (90:0.85) 
                             (210:0.15) -- (210:0.85);

\draw[color=red,shift={(0.86602,1.5)}] (-30:0.15) -- (-30:0.85);

\draw[color=red,shift={(1.732,1)}] (30:0.15) -- (30:0.85);

\draw (0.2,1.8) circle (0.5 pt);
\draw (-0.2,1.8) circle (0.5 pt);
\draw (-0.6,1.8) circle (0.5 pt);

\draw (6.0,0.4) circle (0.5 pt);
\draw (6.4,0.4) circle (0.5 pt);
\draw (6.8,0.4) circle (0.5 pt);

\draw (3.4,-0.7) circle (0.5 pt);
\draw (3.4,-1.1) circle (0.5 pt);
\draw (3.4,-1.5) circle (0.5 pt);
\end{scope}

\end{scope}
\draw (-0.8,3.0) -- (-1.5,3.0) -- (-1.5,-2.0) -- (-0.8,-2.0);
\draw (7.5,3.0) -- (7.5,-2.0);
\draw (15.8,3.0) -- (16.5,3.0) -- (16.5,-2.0) -- (15.8,-2.0);
\draw (-4.8,0.4) node {\Large $3s = S$};
%% 2nd piece  line 2
\begin{scope}[shift={(3,-6)}]

   \begin{scope}[line width=1.0pt]
      \draw[shift={(1*1.732,0)}] (-30:0.15) -- (-30:0.85) (210:0.15) -- (210:0.85);
      \draw[color=red,shift={(1*1.732,0)}] (90:0.15) -- (90:0.85);
      \foreach \i in {2,3}
      {
      \draw[shift={(\i*1.732,0)}] (90:0.15) -- (90:0.85) 
                      (-30:0.15) -- (-30:0.85) (210:0.15) -- (210:0.85);
      }
      \draw[color=red,shift={(0.86602,1.5)}] (90:0.15) -- (90:0.85) 
                      (-30:0.15) -- (-30:0.85) (210:0.15) -- (210:0.85);
      
      \draw[color=red,shift={(1.732,1)}] (30:0.15) -- (30:0.85);
      
      \draw (0.2,1.8) circle (0.5 pt);
      \draw (-0.2,1.8) circle (0.5 pt);
      \draw (-0.6,1.8) circle (0.5 pt);
      
      \draw (6.0,0.4) circle (0.5 pt);
      \draw (6.4,0.4) circle (0.5 pt);
      \draw (6.8,0.4) circle (0.5 pt);
      
      \draw (3.4,-0.7) circle (0.5 pt);
      \draw (3.4,-1.1) circle (0.5 pt);
      \draw (3.4,-1.5) circle (0.5 pt);
      
      %% condition
      \begin{scope}[shift={(10,0)}]
            \draw[shift={(1*1.732,0)}] (-30:0.15) -- (-30:0.85) (210:0.15) -- (210:0.85);
            \draw[color=red,shift={(1*1.732,0)}] (90:0.15) -- (90:0.85);

            \foreach \i in {2,3}
            {
            \draw[shift={(\i*1.732,0)}] (90:0.15) -- (90:0.85) 
                            (-30:0.15) -- (-30:0.85) (210:0.15) -- (210:0.85);
            }
            
            \draw[color=red,shift={(0.86602,1.5)}] (90:0.15) -- (90:0.85) 
                            (-30:0.15) -- (-30:0.85) (210:0.15) -- (210:0.85);
            
            %\draw[shift={(1.732,1)}] (30:0.15) -- (30:0.85);
            
            \draw (0.2,1.8) circle (0.5 pt);
            \draw (-0.2,1.8) circle (0.5 pt);
            \draw (-0.6,1.8) circle (0.5 pt);
            
            \draw (6.0,0.4) circle (0.5 pt);
            \draw (6.4,0.4) circle (0.5 pt);
            \draw (6.8,0.4) circle (0.5 pt);
            
            \draw (3.4,-0.7) circle (0.5 pt);
            \draw (3.4,-1.1) circle (0.5 pt);
            \draw (3.4,-1.5) circle (0.5 pt);
      \end{scope}
   \end{scope}
      
      \draw (-0.3,3.0) -- (-1.1,3.0) -- (-1.1,-2.0) -- (-0.3,-2.0);
      \draw (7.8,3.0) -- (7.8,-2.0);
      \draw (16.8,3.0) -- (17.5,3.0) -- (17.5,-2.0) -- (16.8,-2.0);
   
       \draw (-3,0.4) node {\Large $+ S$};
\end{scope}
\end{tikzpicture}
\end{equation}
The red islands again provide a visual cue.  Following it, we get the bound
\begin{equation}
\begin{tikzpicture}[scale=0.30]
\begin{scope}[style=thick]
\draw (-90:0.15) -- (-90:0.85) (150:0.15) -- (150:0.85);
\draw[shift={(0.5*1.732,0.5)}] (90:0.15) -- (90:0.85) 
                               (-30:0.15) -- (-30:0.85) (210:0.15) -- (210:0.85);
%% condition
\draw[shift={(4,0)}] (-90:0.15) -- (-90:0.85) (30:0.15) -- (30:0.85) (150:0.15) -- (150:0.85);
%brackets
\draw (-0.8,2.0) -- (-1.5,2.0) -- (-1.5,-1.6) -- (-0.8,-1.6);
\draw (2.4,2.0) -- (2.4,-1.6);
\draw (4.8,2.0) -- (5.5,2.0) -- (5.5,-1.6) -- (4.8,-1.6);
\draw (-4.7,0) node {\Large $3s \le S$};
\end{scope}
%2nd piece
\begin{scope}[style=thick,shift={(11,0)}]
\draw[shift={(0,0.5)}] (90:0.15) -- (90:0.85) 
                               (-30:0.15) -- (-30:0.85) (210:0.15) -- (210:0.85);
\draw[shift={(0.5*1.732,0)}] (-90:0.15) -- (-90:0.85) (30:0.15) -- (30:0.85);
%% condition
\begin{scope}[shift={(4,0)}]
\draw[shift={(0,0.5)}] (90:0.15) -- (90:0.85) 
                               (-30:0.15) -- (-30:0.85) (210:0.15) -- (210:0.85);
\draw[shift={(0.5*1.732,0)}] (-90:0.15) -- (-90:0.85);
\end{scope}
%% brackets
\draw (-0.8,2.0) -- (-1.5,2.0) -- (-1.5,-1.6) -- (-0.8,-1.6);
\draw (2.4,2.0) -- (2.4,-1.6);
\draw (5.4,2.0) -- (6.1,2.0) -- (6.1,-1.6) -- (5.4,-1.6);
\draw (-3.5,0) node {\Large $+ S$};
\end{scope}
\end{tikzpicture}
\label{2_step}
\end{equation}

Bounds for a translation-invariant state on the square lattice can be obtained in a
similar way.  We use the one-step bounds
%% simple cross
\begin{equation}
\begin{tikzpicture}[scale=0.30]
\begin{scope}[style=thick]
\draw 
(-90:0.15) -- (-90:0.85) (0:0.15) -- (0:0.85) (90:0.15) -- (90:0.85) (180:0.15) -- (180:0.85);
%% condition
\draw[shift={(4,0)}] (-90:0.15) -- (-90:0.85) (180:0.15) -- (180:0.85);
%brackets
\draw (-0.8,1.6) -- (-1.5,1.6) -- (-1.5,-1.6) -- (-0.8,-1.6);
\draw (2.0,1.6) -- (2.0,-1.6);
\draw (4.8,1.6) -- (5.5,1.6) -- (5.5,-1.6) -- (4.8,-1.6);
\draw (-4.7,0) node {\Large $2s \le S$};
\end{scope}
\end{tikzpicture}
\label{smallcross}
\end{equation}
%% dblcross
and
\begin{equation}
\begin{tikzpicture}[scale=0.30]
\begin{scope}[style=thick]
\draw [shift={(0,-0.5)}] (-90:0.15) -- (-90:0.85) (0:0.15) -- (0:0.85) 
                         (90:0.15) -- (90:0.85) (180:0.15) -- (180:0.85);
\draw [shift={(0,0.5)}] (0:0.15) -- (0:0.85) (90:0.15) -- (90:0.85) (180:0.15) -- (180:0.85);
%% condition
\draw [shift={(4,-0.5)}] (-90:0.15) -- (-90:0.85) (0:0.15) -- (0:0.85) 
                         (90:0.15) -- (90:0.85) (180:0.15) -- (180:0.85);
\draw [shift={(4,0.5)}] (180:0.15) -- (180:0.85);
%brackets
\draw (-0.8,1.6) -- (-1.5,1.6) -- (-1.5,-1.6) -- (-0.8,-1.6);
\draw (2.0,1.6) -- (2.0,-1.6);
\draw (4.8,1.6) -- (5.5,1.6) -- (5.5,-1.6) -- (4.8,-1.6);
\draw (-4.7,0) node {\Large $2s \le S$};
\end{scope}
\end{tikzpicture}
\label{doublecross}
\end{equation}
and the two-step bound
%% box + cross
\begin{equation}
\begin{tikzpicture}[scale=0.30]
\begin{scope}[style=thick]
\draw (-90:0.15) -- (-90:0.85) (0:0.15) -- (0:0.85) 
      (90:0.15) -- (90:0.85) (180:0.15) -- (180:0.85);
%% condition
\draw[shift={(4,0)}] (-90:0.15) -- (-90:0.85) (180:0.15) -- (180:0.85);
%brackets
\draw (-0.8,1.6) -- (-1.5,1.6) -- (-1.5,-1.6) -- (-0.8,-1.6);
\draw (2.0,1.6) -- (2.0,-1.6);
\draw (4.8,1.6) -- (5.5,1.6) -- (5.5,-1.6) -- (4.8,-1.6);
\draw (-4.8,0) node {\Large $2s \le S$};
%2nd piece
\begin{scope}[shift={(10.5,0)}]
\draw[shift={(-0.5,-0.5)}] (0,0.15) -- (0,0.85) (0.15,0) -- (0.85,0) 
                           (1,0.15) -- (1,0.85) (0.15,1) -- (0.85,1);
%% condition
\draw[shift={(3.5,-0.5)}] (0,0.15) -- (0,0.85) (0.15,0) -- (0.85,0) 
                           (1,0.15) -- (1,0.85);
%brackets
\draw (-0.8,1.6) -- (-1.5,1.6) -- (-1.5,-1.6) -- (-0.8,-1.6);
\draw (2.0,1.6) -- (2.0,-1.6);
\draw (4.8,1.6) -- (5.5,1.6) -- (5.5,-1.6) -- (4.8,-1.6);
\draw (-3.2,0) node {\Large $+ S$};
\end{scope}
\end{scope}

\end{tikzpicture}
\label{square_2_step}
\end{equation}

A constrained maximum entropy state on the square lattice with given correlations between diagonal and accross-the-vertex nearest neighbor correlations coincides with a Gibbs state for an Ising model with effective pair interactions for the two types of nearest neighbors of whatever strength is required to reproduce the required correlations.  We have previously\cite{ke08} studied specific pair correlations in such maximum entropy states using Monte Carlo simulation.  Building on the conditioning techniques developed in this Letter, we compute $s(\mathrm{expt}||\mathrm{ME})$, the relative entropy density of an experimental state to the corresponding constrained maximum entropy state.  In the case of two probability measures on the configuration space of a lattice system, the relative entropy of their restrictions to some finite region $\Lambda$ is 
\begin{eqnarray}
S(Q_\Lambda||P_\Lambda) 
&= 
\sum_{{\sigma}_\Lambda} \log_2\left(\frac{Q(\sigma_\Lambda)}{P(\sigma_\Lambda)}\right) \, 
Q(\sigma_\Lambda)
\nonumber \\ 
&= 
\Big\langle \log_2 {Q_\Lambda}\Big\rangle_Q
- \Big\langle \log_2 {P_\Lambda}\Big\rangle_Q.
\label{split}
\end{eqnarray}
The limiting relative entropy density which we want is
\beq
s(Q_\Lambda||P_\Lambda) = \lim_{\Lambda\nearrow\infty} |\Lambda|^{-1} S(Q_\Lambda||P_\Lambda).
\eeq
The logarithm of the probability in Eq. (\ref{split}) can be expanded in terms of conditionals just as was done for the entropy.  For any collection $\{X_1,\ldots, X_N\}$ of random variables (the $m=1$ term being read as an unconditioned probability),
$$
\log_2 P(X_N,\ldots,X_1) = \sum_{m=1}^N \log_2 P(X_m| X_{m-1},\ldots,X_1)
$$
parallels exactly the formula
$$
S(X_N,\ldots,X_1) = \sum_{m=1}^N S(X_m| X_{m-1},\ldots,X_1).
$$
The main difference is that $\log_2 P(\cdot)$ is a random variable, whereas $S(\cdot)$ is a number.  Any of the class of approximations for the conditional entropy densities can now be applied to the conditional probabilities to obtain the relative entropy.  However, we do not get bounds in this way, just ordinary estimates.

If $P_{ME}$ is a maximum entropy state constrained to have expectations of specified observables match their expectations in $P$, then the relative entropy density $s(P||P_{ME})$ ought to equal the difference in the absolute densities.  We believe that the method we have used is superior to this simple subtraction because it suppresses the unwanted effects of fluctuations in the counting of low-probability configurations.  Due to limited experimental data, configurations which would be expected to have only a few occurences may have none at all, which has an anomalously large effect in the subtraction.
%% Here is the endmatter stuff: Supplementary Info, etc.
%% Use \item's to separate, default label is "Acknowledgements"
\noindent{\bf Acknowledgements.}
We acknowledge the financial support from Army Research Office and the National Science Foundation MRSEC program (DMR-0820404) and the National Nanotechnology Infrastructure Network. We are grateful to Prof. Chris Leighton for providing the samples 

%% Put the bibliography here, most people will use BiBTeX in
%% which case the environment below should be replaced with
%% the \bibliography{} command.

\noindent{\bf Author contributions.}
P.S. and V.H.C. conceived the initial idea of this project, X.K., J.L. and D.M.G. performed
the experiments and collected data. C.N. made theoretical contributions.
P.E.L. analyzed data and developed theory.

\noindent{\bf Additional information.}
Supplementary information accompanies this paper at www.nature.com/naturephysics.
Correspondence and requests for materials should be addressed to PEL.

\vfill\eject

%\title{Supplementary Information: Direct Entropy Determination and Application to Artificial Spin Ice}

\section*{Supplementary Information 1: Data Set Statistics}

All data for each combination of lattice spacing and field step size are gathered 
after a single demagnetization run, except for 680 nm and 880 nm with 
$\Delta H_{\mathrm{ext}} =$ 9.6 Oe, which incorporate data from two runs.  
For square lattices, magnetic force microscopy images have size
$10\, \mu$m$ \times 10\, \mu$m, $17\, \mu$m$ \times 17\, \mu$m, 
or $22\, \mu$m$ \times 22\, \mu$m for lattice constant 400 nm, 680 nm
and 880 nm, respectively, in order to have approximately the same number
of islands in each image, although this condition is not important to the results.
A varying number of images are taken at well-separated points on the array.
For honeycomb lattices, a similar procedure was followed.  All images 
are $10\, \mu$m$ \times 10\, \mu$m, except for the runs at 1.6 Oe,
which, by historical accident, 
are $6\, \mu$m$ \times 6\, \mu$m, $8\, \mu$m$ \times 8\, \mu$m, 
$10\, \mu$m$ \times 10\, \mu$m, or $14\, \mu$m$ \times 14\, \mu$m
for the 520 nm, 600 nm, 739 nm and 981 nm lattice, respectively.  

For honeycomb lattices, from the images,
the configuration of every 5-island cluster { \includegraphics[width=10 mm]{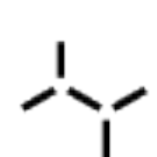} } 
or rotation thereof is recorded as a separated data point and all data
a for given lattice spacing and $\Delta H_{\mathrm{ext}}$ is combined.
The numbers of such data points entering our analysis are as follows:
\medskip

%\begin{table}
\begin{tabular}{c| c c c c}
% \multicolumn{5}{c}{Honeycomb lattice dataset sizes} \\
\hline
% \multirow{2}{*}{Lattice} & \multicolumn{4}{c}{field step (Oe)} \\
\multirow{2}{*}{Lattice} & \multicolumn{4}{c}{field step} \\
% \cline{2-5}
& 1.6 Oe & 3.2 Oe & 9.6 Oe & 16 Oe \\
% & 1.6 & 3.2 & 9.6 & 16 \\
% \hline
\cline{2-5}
% \\[1pt]
520 nm & 831 & 3332 & 1754 & 2836 \\
600 nm & 1620 & 3984 & 3333 & 2291 \\
739 nm & 1964 & 2244 & 2759 & 1770 \\
981 nm & 2096 & 2098 & 3286 & 2317 \\
\hline
\end{tabular}
%\end{table}
\medskip

%\newpage
For square lattices, the 7-island cluster { \includegraphics[width=10 mm]{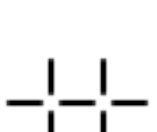} } 
and rotations thereof are used.  The sizes of those data sets are:
\medskip

%\begin{table}
\begin{tabular}{c|cccccc}
%\multicolumn{7}{c}{Square lattice dataset sizes} \\
\hline
%\multirow{2}{*}{Lattice} & \multicolumn{6}{c}{field step (Oe)} \\
\multirow{2}{*}{Lattice} & \multicolumn{6}{c}{field step } \\
%\hline
%\cline{2-7}
& 1.6 Oe & 3.2 Oe & 9.6 Oe & 12.8 Oe & 16 Oe & 32 Oe\\
% & 1.6 & 3.2 & 9.6 & 12.8 & 16 & 32 \\
%\hline
\cline{2-7}
400 nm & 3861 & 2729 & 5523 & 6363 & 4158 & 4699 \\
680 nm & 3848 & 2225 & 1843 & 1793 & 3595 & 1870 \\
880 nm & 3074 & 3196 & 3079 & 3268 & 3870 & 3001 \\
\hline
\end{tabular}
%\end{table}
\medskip

The raw data (MFM scans) for square lattice arrays used here
were the basis of a previously published study\cite{ke08}.

\section*{Supplementary Information 2: Entropies of Honeycomb lattice ideal states}

In this section, we substantiate claims in the Letter about the entropies of 
two ideal states on the honeycomb lattice: quasi-ice and quasi-ice with full sublattice
polarization.
The former is the state obtained by imposing the quasi-ice rule (2-in/1-out or 1-in/2-out)
at every vertex.  The latter is more restricted: vertices on one sublattice must be 
2-in/1-out ($Q=-1$) and those on the second sublattice, 1-in/2-out ($Q=+1$).
This is a broken symmetry state as it entails a choice of which sublattice is which;
for our purposes, we simply pick one.

\subsection*{Quasi-ice}

We will show that the quasi-ice state is equivalent to the ordinary
nearest-neighbor Ising model having spins at the vertices of the honeycomb lattice,
at the special coupling value $\tanh( J/k_B T) = 1/3$.  
On a given graph with periodic boundary conditions, 
the partition functions of the two models differ only by a trivial factor:
quasi-ice degrees of freedom reside on the edges of the graph, those of the Ising model, 
on the vertices.  Quasi-ice is a combinatorial problem, so the logarithm of the
partition function is precisely the entropy we need.  The equivalence equates it
with the {\em free energy\/} of the honeycomb-lattice Ising model at a special temperature,
which is available from the exact solution\cite{baxter-enting,baxter,lavis-bell}.
Alternatively, the expansion developed below can be used directly.

The equivalence is established through a representation of the quasi-ice partition function 
in a high-temperature expansion\cite{simon_lattice_gases, itzykson-drouffe-2}.
Here is a fragment of quasi-ice
\newline
\centerline{ \includegraphics[width=60 mm]{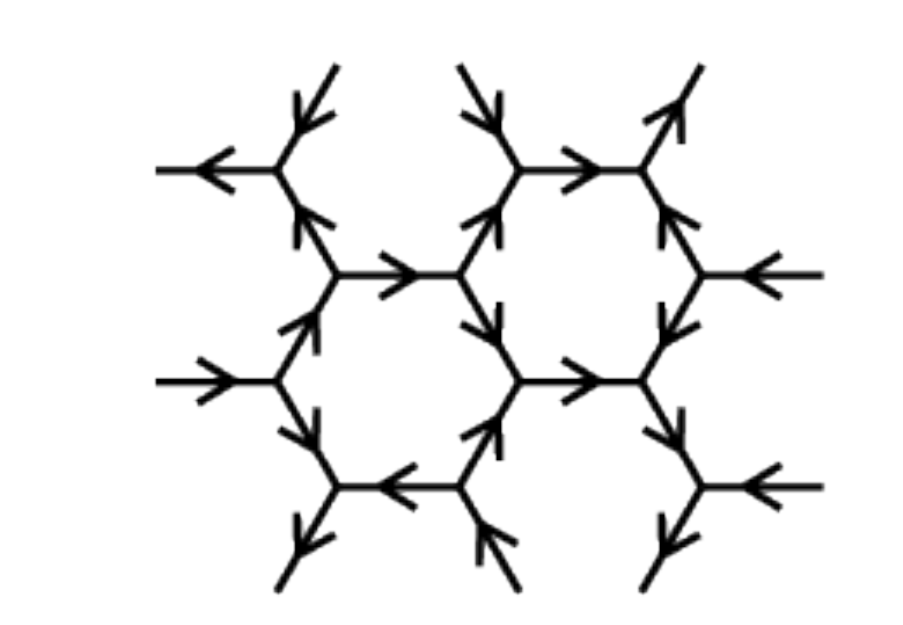} }
\newline
%\begin{figure}
%\epsfxsize 60 mm
%\epsfbox{qifig1.eps}
%\caption{A hexagonal quasi-ice fragment.}
%\label{quasi-ice-fragment}
%\end{figure}
taken from a lattice $\Lambda$ equipped with periodic boundary conditions.
We think of $\Lambda$ as a graph with vertex set $V(\Lambda)$ and edge set $E(\Lambda)$.
The partition function, which simply counts configurations allowed by the quasi-ice rule is
\beq
Z_\Lambda = \sum_{\sigma_\Lambda} \prod_{i\in V(\Lambda)} {\mathcal I}_i(\sigma).
\eeq
The sum is over {\em all} assignments of arrows to the edges (as in the figure),
and the factors ${\mathcal I}_i(\sigma)$ represent the quasi-ice-rule: 
${\mathcal I}_i(\sigma)$ is equal to one if the configuration $\sigma$ on the
edges incident to vertex $i$ satisfy the rule, otherwise it is zero.

To proceed, we introduce extra variables.  {\it Two} ``spins'' are assigned to 
the edge $ij$, $\sigma_i^j$, belonging to vertex $i$ and $\sigma_j^i$, belonging
to vertex $j$.  We can think of them as arrows, or as $\pm 1$-valued objects:
$\sigma_i^j$ is regarded to be $+1$ if the corresponding arrow points away from
vertex $i$.  Then we remove the redundancy by requiring $\sigma_i^j = -\sigma_j^i$,
writing the partition function as
\beq
Z_\Lambda = \sum_{\{ \sigma_i^j\}} 
\prod_{i\in V(\Lambda)} {\mathcal I}_i(\sigma)
\prod_{ij\in E(\Lambda)} \delta(\sigma_i^j,-\sigma_j^i).
\label{redundant}
\eeq
The point of this maneuver is that the Kronecker deltas can be rewritten as
\beq
\delta(\sigma_i^j,-\sigma_j^i) = \frac{1- \sigma_i^j\sigma_j^i}{2},
\eeq
which form allows expansion.
The same trick was used by Nagle\cite{nagle66} for the square lattice
ice model.  The partition function can thus be written as,
\beq
Z = 2^{-|E|} Z^{(0)} 
\left\langle \prod_{ij\in E(\Lambda)} (1 - \sigma_i^j\sigma_j^i ) \right\rangle_0,
\label{quasi-ice-as-expectation}
\eeq
where $\langle \cdot \rangle_0$ denotes expectation in the system of decoupled
vertices, which has partition function 
\beq
Z_\Lambda^{(0)} = \sum_{\{ \sigma_i^j\}} \prod_{i\in V(\Lambda)} {\mathcal I}_i(\sigma).
\eeq

Since the quasi-ice constraint ${\mathcal I}_i$ allows six configurations of spins attached 
the vertex $i$, 
\beq
Z^{(0)} = 6^{|V|} = (6^{2/3})^{|E|}.
\eeq
The prefactor to the expectation in Eq. (\ref{quasi-ice-as-expectation}) is
\beq
%\exp\left[\left(\frac{2}{3}\log 6 - \log 2\right)|E|\right].
e^{(\frac{2}{3}\log 6 - \log 2)|E|}
\label{zeroth-order}
\eeq
The coefficient of $|E|$ in the exponent is $0.50136 = 0.7233$ bit.
This zero-order approximation to the quasi-ice entropy density is already fairly accurate.

The product in the expectation is now expanded as 
\beq
\prod_{ij\in E} (1 - \sigma_i^j\sigma_j^i ) = 
\sum_{G\subset E} \prod_{ij\in G} (-\sigma_i^j\sigma_j^i ).
\label{expansion}
\eeq
The crucial observation at this point is that 
$\langle \sigma_i^j \rangle_0 = 0$ and
$\langle \sigma_i^j \sigma_i^k \sigma_i^l \rangle_0 = 0$ when
$j$, $k$ and $l$ are distinct, because the quasi-ice rule is spin-flip invariant.
Also $\langle \sigma_i^j \sigma_i^k \rangle_0 = -1/3$.
This shows that when the expectation of the expansion in Eq. (\ref{expansion}) is
taken, the term for graph $G$ yields zero unless in has an even number (0 or 2) of edges 
incident on each vertex.  Further, $G$ must consist of simple closed loops, 
no two of which have vertices in common.  This follows from the fact that there
are only three edges of $\Lambda$ incident on each vertex.  If two belong to $G$,
the third cannot, and the first two must be part of a simple closed loop.
For each vertex in a loop, we get a factor $\langle \sigma_i^j \sigma_i^k \rangle_0 = -1/3$,
and for each edge, a factor $-1$ [from Eq. (\ref{expansion})].  Since a closed loop $\gamma$
has an equal number of edges and vertices, the minus signs cancel and its weight is
simply $(1/3)^{|\gamma|}$.
Thus,
\beq
\left\langle \prod_{ij\in E(\Lambda)} (1 - \sigma_i^j\sigma_j^i ) \right\rangle_0 =
%\sum_{\{\gamma_\alpha\}\,  \atop \mbox{\small disjoint}} 
\sum_{ \mbox{\small disjoint} \{\gamma_\alpha\} } 
\prod_\alpha \left(\frac{1}{3}\right)^{|\gamma_\alpha|}.
\label{loop-expansion}
\eeq
By comparing to the high-temperature expansion of the honeycomb lattice Ising model, 
$ ({3^{1/3}}/{2^{1/2}})^{|E|} Z_{\Lambda} $
is seen to be equal to the Ising model partition function at $\tanh J = 1/3$.
From an exact solution for the latter model (references above), 
$s=0.724$ bit/island.  It is also practical to apply polymer expansion 
techniques to the right-hand side of Eq. (\ref{loop-expansion}).

\subsection*{Vertex-charge-ordered quasi-ice}

If the quasi-ice constraint is strengthened to require that all vertices on one sublattice
are 2-in/1-out ($Q=-1$) and all on the second sublattice are 1-in/2-out ($Q=+1$), 
the vertex-charge-ordered quasi-ice state is obtained.  
We show how to match the configurations of this state 
one-to-one with those of the triangular lattice Ising antiferromagnet at zero
temperature.  Since an exact solution\cite{wannier50,houtappel50,lavis-bell} is available for the latter,
the problem is again solved.  The equivalence in this case is a simple observation.
Each vertex on the first sublattice has precisely one {\it out} island, and that
island is the only one pointing {\it in} to the vertex at its other end.
Thus, each vertex-charge-ordered quasi-ice configuration corresponds
to one {\it dimer covering} of the lattice: a selection of edges such that each vertex is
touched by exactly one. For example,
\newline
\centerline{ \includegraphics[width=60 mm]{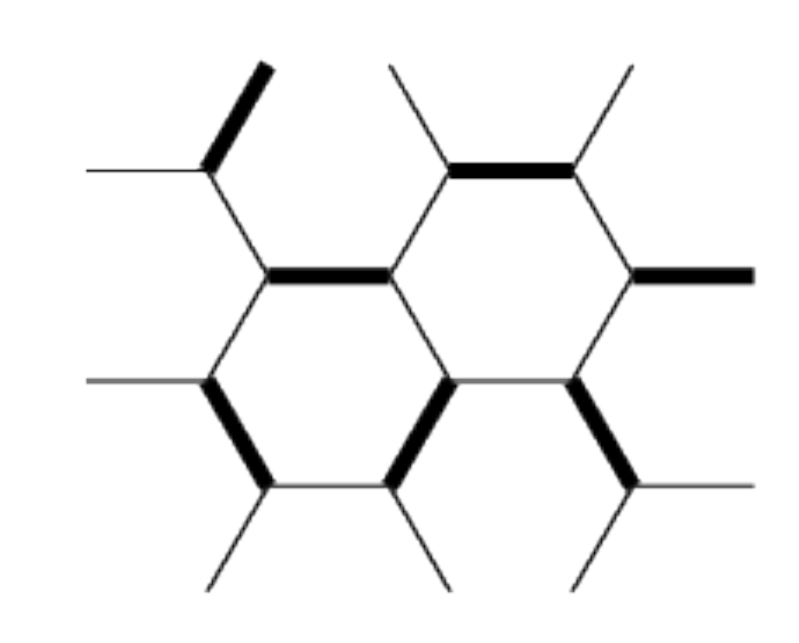} }
\newline
Passing to the dual lattice, which is a triangular lattice, we get a picture like this
\newline
\centerline{ \includegraphics[width=60 mm]{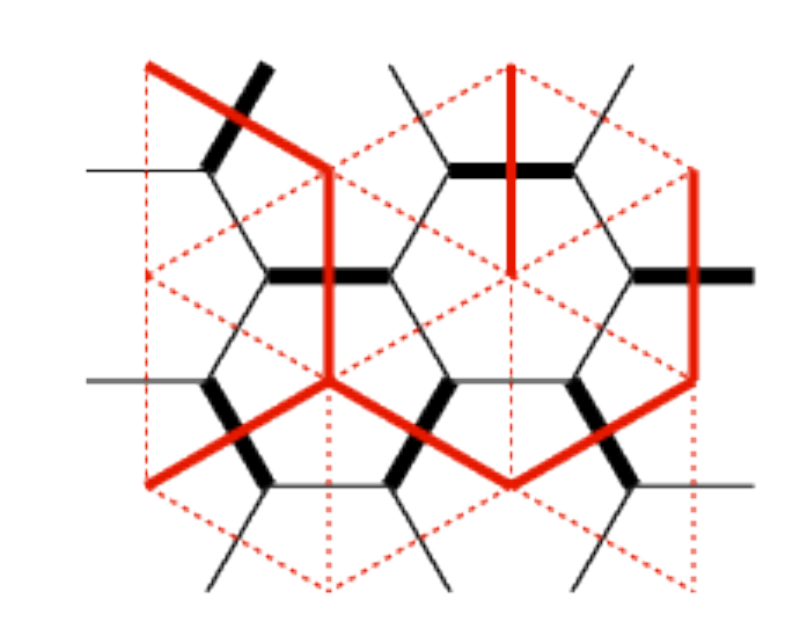} }
\newline
in which one edge of each triangle is selected.  Now identifying these as the frustrated
bonds of an Ising model on the triangular lattice ($\sigma_i\sigma_j = +1$),
it is clear that the dimer coverings are in one-to-one correspondence with ground
state configurations of the triangular lattice Ising antiferromagnet.
Transcribing exact results for that model\cite{wannier50,houtappel50,lavis-bell},
$s = 0.155$ bit/island.

The vertex-charge-ordered quasi-ice state is also equivalent to a 5-vertex model
in which one of the vertices of the usual 6-vertex model is forbidden (see figure below).
Wu\cite{wu68} first noted the equivalence of the honeycomb lattice dimer problem with
a 5-vertex model under the name ``modified KDP model.''
Bl\"ote and Hilhorst\cite{blote82} noticed the connection with the triangular lattice
Ising antiferromagnet.  The 5-vertex model is of interest as a model of the terrace-ledge-kink
picture of surface growth\cite{garrod90}.

\begin{figure}
\includegraphics[width=60 mm]{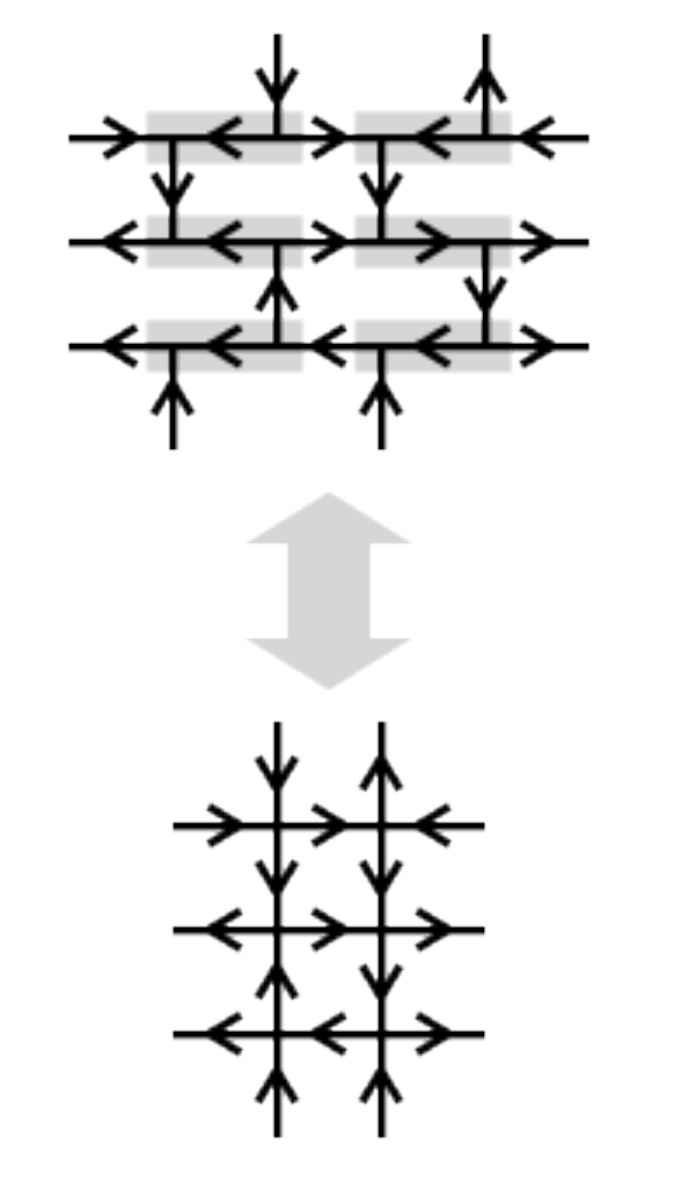} 
\caption{Equivalence between vertex-charge-ordered quasi-ice and a 5-vertex model.}
\label{5-vtx-fig}
\end{figure}

We indicate how the connection is made.  First, deform the angles in the honeycomb to
make a brickwork (ladder) lattice as at the top of Figure \ref{5-vtx-fig}. 
Vertices are paired up according to the shaded blocks.  Without loss of generality,
assume that the sublattice symmetry is broken such that the left-hand vertices
in each block are 2-in/1-out and the right-hand ones are 1-in/2-out.  
It is easy to see that the arrows on the edges inside the blocks are fully determined by 
the others.  So, the blocks may be contracted to vertices on a square lattice as at the
bottom, and the vertices can be uniquely re-expanded to reconstruct the original lattice.
The vertices which occur (with equal weight) are five of the six vertices of the 6-vertex 
square ice model\cite{baxter}; one is forbidden,
\newline
\centerline{ \includegraphics[width=85 mm]{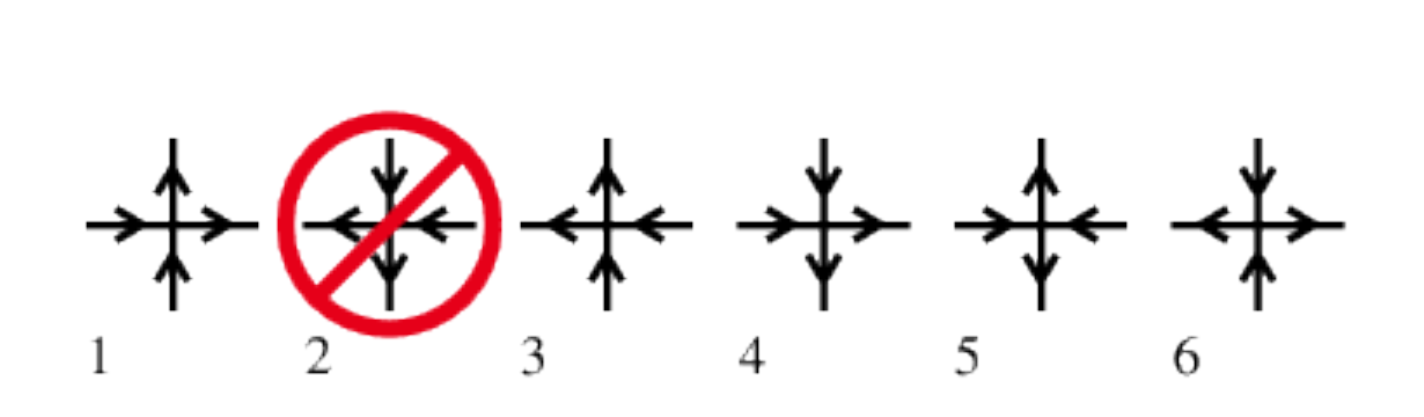} }
\newline
because it has two outgoing (ingoing) arrows for the left-hand (right-hand) vertex 
of a block.

\section*{Supplementary Information 3: macrostate mixtures}

Suppose the macrostate produced by the experimental protocol was nonunique,
for example, if the array was left with macroscopic magnetization pointing in 
a random direction, or if the island interactions led to a spontaneous symmetry breaking.   
Then, sampling over the course of multiple runs would be tantamount to 
sampling from a mixture of macrostates, corrupting the entropy measurement
with a spurious entropy of the mixture.  
It is interesting to see how much our technique is affected by this problem. 
Labelling distinct macrostates by $\theta$, assume for simplicity that they all have
the same entropy density, as would hold in case of a broken symmetry.
Then, what we want is the entropy conditioned on $\theta$. 
Apropos the straightforward method, observe that
$S(\sigma)-S(\sigma_\Lambda|\theta) = S(\theta)-S(\theta|\sigma_\Lambda)$,
so that the spurious entropy counted for a cluster is the average information 
that cluster provides about the macrostate parameter. 
On the other hand, for the conditioning technique, 
$S(\sigma_\Lambda|\sigma_\Gamma)-S(\sigma_\Lambda|\sigma_\Gamma,\theta) 
=S(\theta|\sigma_\Lambda,\sigma_\Gamma)-S(\theta|\sigma_\Gamma)$, so that
the spurious entropy count is only the {\em additional\/} information about
$\theta$ brought by $\sigma_\Lambda$, over what is already provided by $\sigma_\Gamma$.   
In general, this would seems to be another superiority of our technique
compared to the straightforward one.

The relation of this to our data analysis has two sides.
First, for most combinations of lattice spacing and $\Delta H_{\mathrm{ext}}$,
our data come from a single run anyway.  However, there is ample 
evidence that there is no overt cooperative symmetry breaking; the only
sort of multiple macrostates would be a residual magnetization.
We have symmetrized our raw data over discrete lattice symmetries,
as a way to reduce sampling error under the assumption that the underlying
state respects those symmetries.    
There are indications that the largest step size (32 Oe) may result in a 
slight residual macroscopic magnetization, in which case the symmetrization
procedure may be questioned.  Even supposing the residual magnetization to
be present, however, the symmetrization only reduces it by a small discrete
symmetry and therefore leaves most of its effects intact. 

%%
%% TABLES
%%
%% If there are any tables, put them here.
%%

%\bibliography{ent}

\end{document}